\documentclass[aps,prd,onecolumn,superscriptaddress,8pt]{revtex4}
\usepackage[dvips,final]{graphicx}
\usepackage{amsmath}
\usepackage{amssymb}
\usepackage{url}
\usepackage{subfigure}
\usepackage{textcomp}
\usepackage{graphicx}
\usepackage{bm}
\usepackage{dcolumn}
\usepackage[usenames]{color}

\newcommand{\be}{\begin{equation}}
\newcommand{\ee}{\end{equation}}
\newcommand{\bea}{\begin{eqnarray}}
\newcommand{\eea}{\end{eqnarray}}

\newcommand{\bse}{\begin{subequations}}
\newcommand{\ese}{\end{subequations}}
\newcommand{\bi}{\begin{itemize}}
\newcommand{\ei}{\end{itemize}}

\def\beq{\begin{equation}}
\def\eeq{\end{equation}}
\def\bea{\begin{eqnarray}}
\def\eea{\end{eqnarray}}

\def\bal#1\eal{\begin{align}#1\end{align}}

\begin{document}

\title{\Large Dipole Modulation in Tensor Modes: Signatures in CMB Polarization}

\author{Moslem Zarei}
\email[]{m.zarei@cc.iut.ac.ir}

\affiliation{Department of Physics, Isfahan University of
Technology, Isfahan 84156-83111, Iran}

\affiliation{ School of Astronomy, Institute for Research in Fundamental
Sciences (IPM), P. O. Box 19395-5531, Tehran, Iran}

\date{\today}

\begin{abstract}
In this work we consider a dipole asymmetry in tensor modes and study the effects of this asymmetry on the angular power spectra of CMB. We derive analytical expressions for the $C_{l}^{TT}$ and $C_{l}^{BB}$ in the presence of such dipole modulation in tensor modes for $l<100$. We also discuss on the amplitude of modulation term and show that the $C_{l}^{BB}$ is considerably modified due to this term.

\end{abstract}

\date{\today}

\maketitle

\section{ Introduction}

The anomalies such as power asymmetry in the CMB map reported by Planck \cite{Ade:2013nlj} and WMAP \cite{Bennett:2012zja,Hinshaw:2012fq} teams have gained a great deal of attention to the anisotropic inflationary models in recent years \cite{Watanabe:2009ct,Emami:2010rm,Dulaney:2010sq,Gumrukcuoglu:2010yc,Watanabe:2010fh,Bartolo:2012sd,Funakoshi:2012ym,
Yamamoto:2012sq,Shiraishi:2013vja,Abolhasani:2013zya,Abolhasani:2013bpa,Chen:2013eaa,Emami:2014tpa,Kim:2013gka,Eriksen:2003db,Hansen:2004vq,Notari:2013iva,Akrami:2014eta}. The planck team has revisited the phenomenological well studied model of dipole modulation \cite{Hansen:2008ym,Eriksen:2007pc}, originally proposed by Gordon et al. \cite{Gordon:2006ag} and parameterizing as
\beq
\delta T(\mathbf{n})=\delta T_{\textrm{iso}}(\mathbf{n})(1+A\,\textbf{n}\cdot \textbf{p})~,
\eeq
where $\delta T_{\textrm{iso}}(\mathbf{n})$ is the isotropic temperature fluctuations observed in a
direction $\textbf{n}$, $\mathbf{p}$  is the preferred direction and $A$ is the dimensionless amplitude of the dipole asymmetry. The planck team has found a dipole asymmetry at the direction $(227, -15) \pm 19$ in galactic coordinates for large angular
scales with the amplitude $A= 0.078^{+0.020}_{-0.021}$ at the $3.5\,\sigma$ significance level \cite{Ade:2013nlj}. Such observations have encouraged several people to study the models which predict small primordial anisotropy in power spectrum of perturbations.
 In the standard cosmological models the requirements of isotropy and homogeneity can be regarded as the invariance of space under rotation and translation at sufficiently large scales. Then the FRW metric is manifestly written to be invariant under space translations and rotations. The assumption of isotropy also implies that the energy-momentum tensor has to be diagonal with the equal spatial components. At the perturbation level the two point correlation function for curvature perturbations calculated at two different positions $\mathbf{x}$ and $\mathbf{x}'$ is given as a function of $\mathbf{x}-\mathbf{x}'$ due to translation invariance. On the other hand the rotational invariance means that the two point correlation function is given as a function of $|\mathbf{x}-\mathbf{x}'|$ or equivalently in the momentum space the power spectrum is not dependent on the direction of momentum. In order to generate the anisotropy we have to break the rotation invariance. A primordial vector field aligned in a preferred direction can break the SO(3) symmetry group down to the SO(2). The anisotropic inflationary models with vector field impurity has been studied with great interest during recent years \cite{Namba,Emami:2013bk,Soda:2012zm,Himmetoglu}. In these models the primordial vector fields violating the rotational symmetry at early times, leave anisotropic effects on cosmological correlation functions. One can use the remaining SO(2) symmetry to simplify the perturbation calculations and derive a primordial power spectrum which explicitly depends on momentum direction \cite{Namba,Emami:2013bk,Soda:2012zm,Himmetoglu}.

Another approach is the generation of dipole asymmetry in the power spectrum using the long wavelength super-horizon scalar modes \cite{Lyth:2013vha}. It is shown that the local non-Gaussianity in squeezed limit when one mode is super-horizon leads to power spectrum with a dipole asymmetry correction term. Hence, the amplitude of anisotropy is controlled by the local non-Gaussianity parameter $f_{NL}$ \cite{Lyth:2013vha} (see \cite{
Namjoo:2013fka,Liddle:2013czu,Mazumdar:2013yta,Abolhasani:2013vaa,Chang:2013lxa,Firouzjahi:2014mwa,Namjoo:2014nra,Baghram:2014nha,Mirbabayi:2014hda} for recent developments). The dipole asymmetry in the power spectrum is translated to the modulation in the curvature perturbation $\zeta_{k}$ whereas for large scales it is equivalent to the dipole modulation in the CMB temperature anisotropy, $\Delta T(\mathbf{n})$, studied by Planck and WMAP teams \cite{Ade:2013nlj,Bennett:2012zja,Hinshaw:2012fq}. Following the same logic one can show that the super-horizon scalar modes can also modulate the power spectrum of tensor perturbations though with smaller amplitude \cite{Abolhasani:2013vaa}.

In this paper, we consider the modulation in the amplitude of tensor modes originally applied to scalar perturbations in \cite{Erickcek:2008jp}. The dipole modulation in the tensor modes is the implementation of a preferred direction in the amplitude which makes changes in the value of amplitude from one side of the sky to the other side. It is worth to note that the dipole asymmetry is produce by a spatially-dependent tensor power spectrum and a momentum
direction-dependent power spectrum cannot produce a CMB dipole
asymmetry. Here we study the effects of such modulation on the CMB correlations on large angular scales (see also \cite{Namjoo:2014pqa} for the same idea). Because tensor and scalar modes do not interfere, we can deal with the contribution of scalar and tensor modes to CMB angular power spectrum separately. Hence, we write $C^{XY}_{l}=C^{XY(\zeta)}_{l}+C^{XY(t)}_{l}$ where we are including labels t and $\zeta$ to distinguish the angular power spectrum due to tensor modes, $C^{XY(t)}_{l}$, from the curvature perturbations $C^{XY(\zeta)}_{l}$. The spectrum $C^{TT(t)}_{l}$ decays rapidly for $l>50$. For $l\sim 10$ where the contribution of Sachs-Wolf effect is dominant we have $C^{TT(t)}_{l}/C^{TT(\zeta)}_{l}\sim r$ with $r$ denoting the tensor-to-scalar ratio. The E-mode correlation $C^{EE(t)}_{l}$ has a maximum at $l\sim 100$ and decays after $l>100$ \cite{Flauger:2007es}. For this spectrum we have $C^{EE(t)}_{l}/C^{EE(\zeta)}_{l}\sim 0.1\,r$. As well as for the TE cross correlation we find $C^{TE(t)}_{l}/C^{TE(\zeta)}_{l}\sim 0.1\,r$. Therefore, we expect that the modulation in tensor modes leads to larger imprints on $C^{TT(t)}_{l}$. However, the contribution of tensor modes is subdominant in $C^{TT}_{l}$. Consequently, we do not expect to see a significant effect on $C^{TT}_{l}$ due to the modulation in tensor modes . On the other hand, the B-mode polarization is directly related to the amplitude of tensor modes. Hence, $C^{BB}_{l}$ will be more sensitive to the dipole modulation in tensor modes. In this work we first analytically calculate the $C^{BB}_{l}$ and show that it is in good agreement with results of CAMB \cite{Lewis:1999bs} for $10<l<100$. Then we derive the modulated $C^{BB}_{l}$ and investigate the effects of dipole modulation on the $C^{BB}_{l}$. The tensor modulation would not produce an asymmetry in large scale structure. This is consistent with the null detection of a dipole modulation in large scales \cite{Hirata,Fernandez-Cobos:2013fda}. We also anticipate that the tensor anomalies considered here must not also produce strong intrinsic asymmetry on small angular scales \cite{Flender:2013jja}.

The paper is organized as follows: In the next section we first obtain the transfer function for the tensor modes. In section III we discuss the effects of modulation in tensor modes on the $C_l^{TT}$. Finally in section IV we compute the $C_l^{BB}$ in the presence of dipole asymmetry in the tensor modes.

\section{Transfer function of tensor modes}

We write down the perturbed FRW metric in the following form
\beq
ds^{2}=a^{2}(\eta)[-(1+2\Phi)d\eta^{2}-2B_id\eta dx^{i}+(\delta_{ij}+h_{ij})dx^{i}dx^{j}]~,
\eeq
where $\eta$ is the conformal time, $a(\eta)$ is the scale factor and $\Phi$, $B_i$ and $h_{ij}$ are the scalar, vector and tensor perturbations of the metric.
The tensor perturbations are characterized by the transverse traceless tensor $h^{TT}_{ij}$ and using the Einstein equations is governed by the following equation
\beq
h_{ij}^{''TT}(\eta,\mathbf{x})+2\,\frac{a'}{a}\,h_{ij}^{'TT}(\eta,\mathbf{x})-\partial_{i}\partial^{i}h_{ij}^{TT}(\eta,\mathbf{x})=0~,
\eeq
where the prime denotes derivative with respect to conformal time. We apply the decomposition technique to the tensor modes and write $h_{ij}^{TT}(\eta,\mathbf{x})=h_{ij}^{TT}(\eta,\mathbf{k})\,e^{-i\mathbf{k}\cdot\mathbf{x}}$ where $\mathbf{x}=(\eta_0-\eta)\,\mathbf{n}$ will be the distance from the last scattering surface and $\mathbf{n}$ is the direction of photon propagation. In order to calculate the CMB power spectra it is convenient to rotate the coordinate system so that the wave vector $\mathbf{k}$ is aligned along z axis. Hence one can write $\mathbf{k}\cdot\mathbf{n}=k\cos\theta$. The tensor perturbations $h_{ij}^{TT}(\eta,\mathbf{k})$ are separated into the fourier modes of two polarization states,
\bea
h_{ij}^{TT}(\eta,\mathbf{k})=\sum_{A=+,\times}e_{ij}^{(A)}H(k,\eta) h_{(i)}^{(A)}(\mathbf{k})~,
\eea
where $h_{(i)}^{(A)}$ is the primordial gravity wave amplitude, $H(k,\eta)$ is the transfer function and $e_{ij}^{(+)}$ and $e_{ij}^{(\times)}$ are the two symmetric transverse traceless basis tensors. The transfer function $H(k,\eta)$ is governed by the following equation
\beq
H''+2\,\frac{a'}{a}\,H'+k^{2}H=0~, \label{Heq1}
\eeq
where we have ignored the source term due to neutrino anisotropic stress \cite{Weinberg:2003ur}. Here and elsewhere we do not include the neutrino perturbations in our calculations. One can show that for a mixture of radiation and matter fluid the Friedmann equation gives the scale factor as \cite{Giovannini:2007xh}
\beq
a(\eta)=a_{\textrm{eq}}\left[\left(\frac{\eta}{\eta_1}\right)^{2}+2\frac{\eta}{\eta_1}\right]~,\label{sf}
\eeq
where $a_{\textrm{eq}}$ is the value of scale factor at the time of equality and $\eta_1\simeq 78.8\,\Omega^{-1}_{m}$ with the parameter $\Omega_{m}$ denoting the current abundance of matter. The equation \eqref{Heq1} can be solved numerically using the scale factor \eqref{sf}. The results are presented in Fig. 1. As we can see in Fig. 1(a), for those modes with $k\gg k_{\textrm{eq}}(\approx 0.01 \textrm{Mpc}^{-1})$, the numerical results are in good agreement with the analytic solution $H(k,\eta)=\sin(k\eta)/k\eta$ in radiation dominated era. As well as for those long wavelength modes which enter the horizon after equality the numerical solution is in agreement with the analytic solution $3j_{1}(k\eta)/k\eta$ where $j_{1}(x)\equiv (\sin x-x\cos x)/x^2$ is the spherical Bessel function.
For reasons that will become clear later on when we will calculate the $C^{BB}_{l}$, we are interested in the modes which enter the horizon at the time of recombination $\eta_{r}\simeq 288\,\textrm{Mpc}$. Usually, at this time the analytical solution $3j_{1}(k\eta)/k\eta$ is approximated as the transfer function \cite{Turner:1993vb,Gorbunov:2011zzc,Pritchard:2004qp}. Interestingly, as we can see in Fig. 1(b) the numerical solution of equation \eqref{Heq1} for transfer function has a closer agrement with the analytical result $\sin(k\eta)/k\eta$ at $\eta=\eta_r$. Moreover, our later calculations in section IV deriving $C^{BB}_{l}$, suggest that the $\sin(k\eta)/k\eta$ solution is an appropriate transfer function at the time $\eta=\eta_{r}$.

\begin{figure}
    \centering
        \subfigure[  ]
    {
        \includegraphics[width=3.4in]{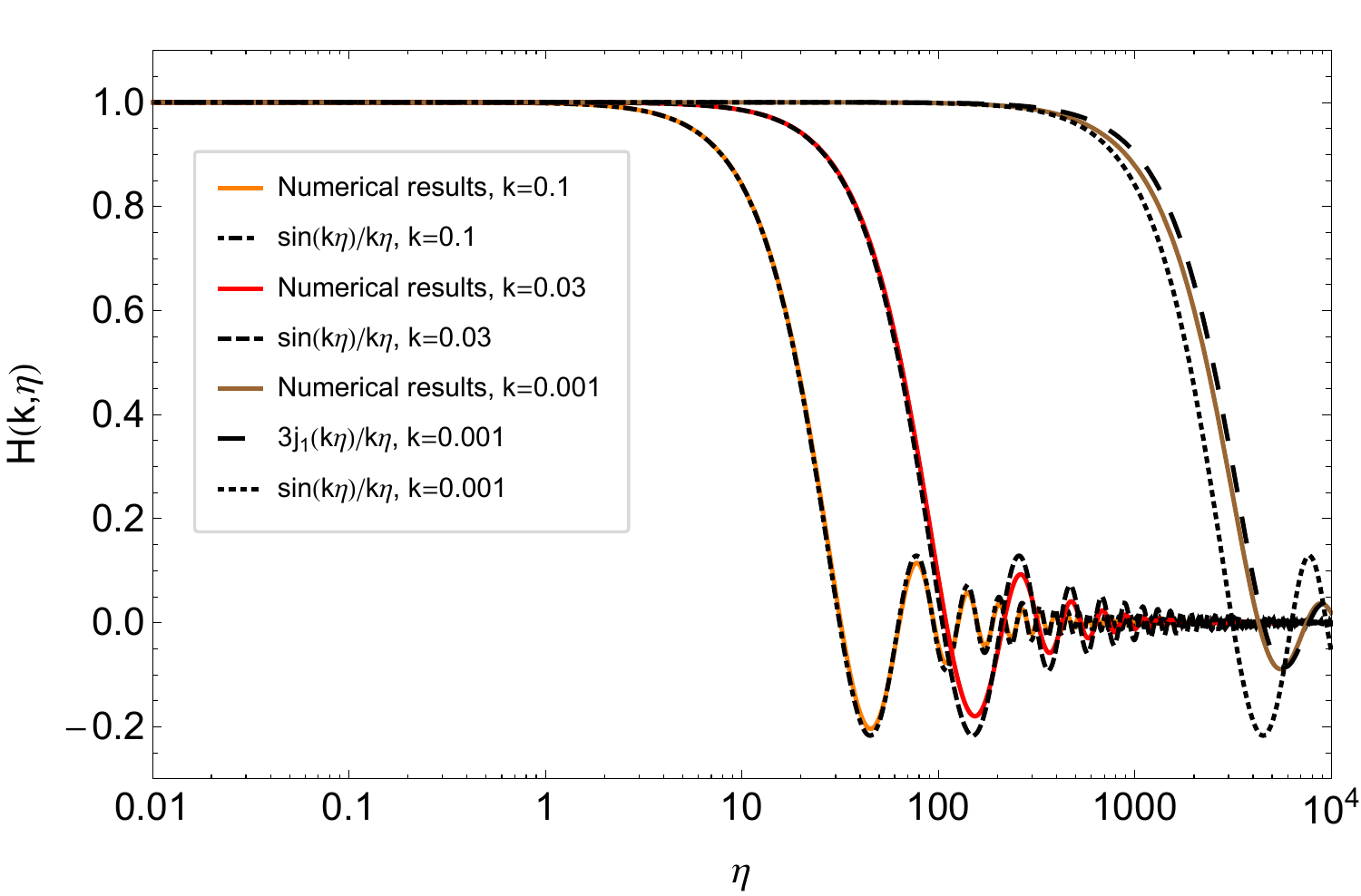}
        \label{fig:second_sub}
    }
       \subfigure[]
    {
        \includegraphics[width=3.4in]{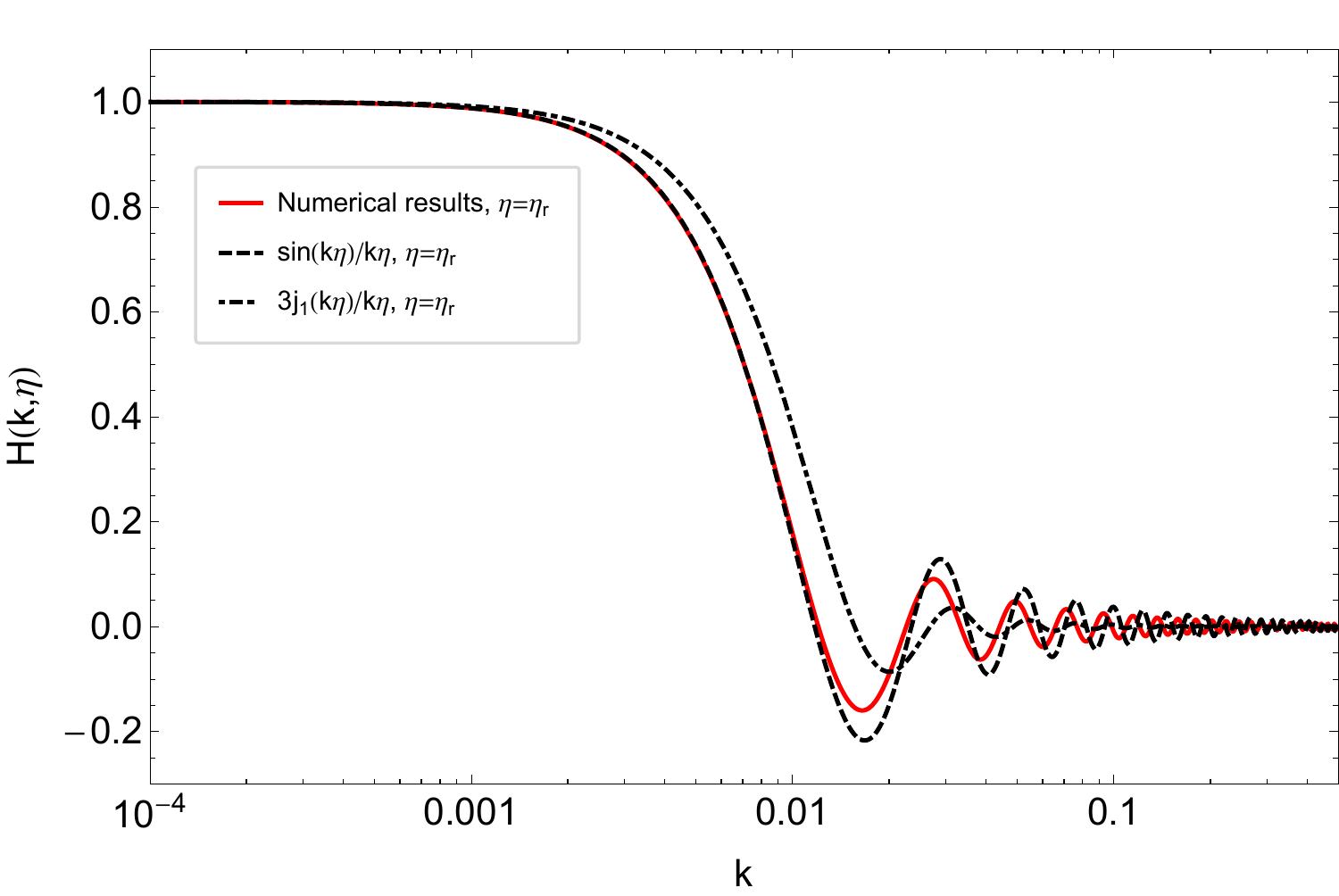}
        \label{fig:third_sub}
    }
    \caption{Comparison of numerical results for the transfer function with the analytic solutions.}
    \label{fig:sample_subfigures}
\end{figure}

A gravitational wave carrier can be modulated similar to what occurs in wave mechanics. In a simplified picture the modulation may be due to a superhorizon long wave tensor mode. The long wavelength mode can change the amplitude of gravity wave in an especial direction from one side of the sky to the other. Adopting a dipole asymmetry term to the position dependent part of the tensor mode we obtain
\bea
h_{ij}^{TT}(\eta,\mathbf{x})=\sum_{A=+,\times}e_{ij}^{(A)}H(k,\eta) h_{(i)}^{(A)}(\mathbf{k})\,e^{-i\mathbf{k}\cdot\mathbf{x}}[\,1+(\mathbf{k}\cdot \mathbf{x})_{_{\textrm{lss}}}]~,\label{modtensor1}
\eea
where $\mathbf{x}_{_{\textrm{lss}}}$ is assumed to be preferred direction at the last scattering surface (lss). Similar to what is done in scalar perturbation case, a spatially-dependent dipole asymmetry power spectrum of the tensor perturbation can be represented in the form
\beq
\mathcal{P}_{t}^{1/2}=\left[1+A_{t}\,\mathbf{n}_{k}\cdot \hat{\mathbf{x}}_{_{\textrm{lss}}}\right]\mathcal{P}_{\textrm{iso},t}^{1/2}
\label{anisop1}
\eeq
 where $A_{t}= (k\,x)_{\textrm{lss}}$ and $\mathcal{P}_{\textrm{iso},t}$ is the isotropic tensor power spectrum that is set by the amplitude of scalar amplitude $A_s$ as $\mathcal{P}_{\textrm{iso},t}=r\,A_s$. As a result, an observer sees a dipole asymmetry in the direction $\hat{\mathbf{x}}_{_{\textrm{lss}}}$ corresponding to the amplitude $A_{t}$. In order to track the impact of such dipole asymmetry on the CMB temperature and polarization power spectra we first assume that $\mathbf{n}_k$ is directed along the $z$ direction and $\mathbf{n}_{k}\cdot \hat{\mathbf{x}}_{_{\textrm{lss}}}=\cos\theta$. Then we use \eqref{modtensor1} and
  analytically recalculate the tensor part of CMB multipoles by considering the following replacement
\beq
e^{-ik(\eta_0-\eta)\cos\theta}\rightarrow e^{-ik(\eta_0-\eta)\cos\theta}[\,1+A_{t}\cos\theta]
\eeq
where the angular integration over the $\theta$ will contribute corrections to the CMB power spectra.

\section{Dipole modulation in CMB temperature power spectrum }

In the absence of the modulation, the contribution of tensor perturbations to the CMB temperature anisotropy is parameterized as \cite{Gorbunov:2011zzc,Weinberg:2008zzc}
\beq
\Theta^{t}(\mathbf{n})=  \frac{1}{2}\int_{\eta_r}^{\eta_0}d\eta\, n_ih^{'TT}_{ij}(\eta,-\mathbf{x})n_j~.\label{Theta}
\eeq
Here $\Theta^{t}$ is the brightness function where the superscript t indicates that the CMB temperature anisotropy is due to tensor modes. The $n_i$ and $n_j$ coefficients are also the unit vectors along the photon momentum and the integral in Eq.\eqref{Theta} is computed along the photon trajectory from the the recombination time, $\eta_r$ to the present time $\eta_0$. In Fourier space the $\Theta^{t}(\mathbf{n})$ is represented in the following form
\bea
\Theta^{t}(\mathbf{n})&=&\frac{1}{2}\int d^{3}k \int_{\eta_r}^{\eta_0}d\eta\,\frac{\partial H}{\partial\eta}\,e^{i(\eta_0-\eta)\mathbf{k}\cdot\mathbf{n}}\sum_{A}n_{i}n_{j}e^{(A)}_{ij}h_{(i)}^{(A)}(\mathbf{k})\nonumber \\&&
\!\!\!\!\!\!\!\!=\frac{1}{2}\int d^{3}k \int_{\eta_r}^{\eta_0}d\eta\,\frac{\partial H}{\partial\eta}\,\sum_{A}n_{i}n_{j}e^{(A)}_{ij}h_{(i)}^{(A)}(\mathbf{k})\sum_{l'=0}^{\infty}(2l'+1)i^{l'}P_{l'}(\cos\theta)j_{l'}\left[(\eta_{0}-\eta)k\right]~,\label{thetat1}
\eea
where we have made use of the expansion of the exponential in terms of Legendre polynomials $P_l$
\beq
e^{i(\eta_0-\eta)\mathbf{k}\cdot\mathbf{n}}=\sum_{l'=0}^{\infty}(2l'+1)i^{l'}P_{l'}(\cos\theta)j_{l'}\left[(\eta_{0}-\eta)k\right]~.\label{planew}
\eeq
One can expand the brightness function $\Theta^{t}$ into multipoles $a^{t}_{lm}$
\beq
\Theta^{t}(\mathbf{n})=\sum_{l=2}^{\infty}\sum_{m=-l}^{m=l}a^{t}_{lm}Y_{lm}(\mathbf{n})~,
\eeq
with $Y_{lm}(\mathbf{n})$ the spherical harmonic functions. Using the orthogonality of spherical harmonics and the convolutions $n_{i}n_{j}e^{(+)}_{ij}=\sin\theta\cos 2\phi$ and $n_{i}n_{j}e^{(\times)}_{ij}=\sin\theta\sin 2\phi$ in spherical frame $(\theta,\phi)$ we arrive at
\bea
&&a_{lm}^{t}=\frac{1}{4}\int d\mathbf{n}Y_{lm}^{\ast}(\mathbf{n})\sin^2\theta[e^{2i\phi}(h_{(i)}^{(+)}-ih_{(i)}^{(\times)})+e^{-2i\phi}(h_{(i)}^{(+)}+ih_{(i)}^{(\times)})]\nonumber \\
&&
\:\:\:\:\:\:\:\:\:\:\:\:\times\sum_{l'=0}^{\infty}(2l'+1)i^{l'}P_{l'}(\mathbf{n}\cdot\mathbf{n}_{k})\int_{\eta_r}^{\eta_0}d\eta\frac{\partial H}{\partial\eta}j_{l'}[(\eta_0-\eta)k]
\eea
To reduce this expression we use the recursion and orthogonality relations for Legendre polynomials
\beq
(1-x^2)\frac{dP_n}{dx}=nP_{n-1}-nxP_n~,\:\:\:\:\:\:
(2n+1)xP_{n}=nP_{n-1}+(n+1)P_{n+1}
\:\:\:\:\:\textrm{and} \:\:\:\:\:
\int_{-1}^{1}d\,x P_n(x)P_m(x)=\frac{2\delta_{nm}}{2n+1}~,\label{legendre}
\eeq
and after some straightforward calculations we find the multipoles as
\beq
a_{l\pm 2}^{t}=\left[h_{(i)}^{(+)}\mp ih_{(i)}^{(\times)}\right]\pi i^{l}\sqrt{\frac{2l+1}{4\pi}\frac{(l+2)!}{(l-2)!}}\int_{\eta_r}^{\eta_0}d\eta\frac{\partial H}{\partial\eta}
\left[\frac{j_{l}[(\eta_0-\eta)k]}{(\eta_0-\eta)^2k^2}\right]~.
\eeq
where the $m=\pm 2$ is appeared as a result of integration over the azimuthal angle $\phi$. After calculating the $a_{l\pm 2}^{t}$ coefficients, one can also take into account the angular power spectrum $C^{TT}_l$. Here we must distinguish between the anisotropies from scalar and tensor modes. The total angular power spectrum in general is written as $C^{TT}_l=C^{TT(\zeta)}_l+C^{TT(t)}_l$. The spectrum due to tensors, $C^{TT(t)}_l$, is given in the following manner
\bea
C^{TT(t)}_l&&=\frac{1}{2l+1}\int d^{3}k\sum_{m=-l}^{l}\left<\mid a^{t}_{lm}(\mathbf{k})\mid^{2}\right>\nonumber \\ &&=\frac{1}{2l+1}\int d^{3}k\left<\mid a^{t}_{l2}(\mathbf{k})\mid^{2}+\mid a^{t}_{l-2}(\mathbf{k})\mid^{2}\right>~.\label{cltt}
\eea
Using the two point correlation function of the primordial tensor perturbation $h^{A}_{(i)}$ with polarization $A=+, \,\times$ one can write
\beq
\left<\mid h_{(i)}^{(+)}\mp ih_{(i)}^{(\times)}\mid^{2}\right>=\frac{1}{8\pi k^{3}}\,\mathcal{P}_{\textrm{iso},t}~. \label{2pow}
\eeq
 After changing the variables of integration from $k\eta_0$ to $u$ and $\eta/\eta_0$ to $\xi$ and using the fact that
  $H(k,\eta)=3j_1(k\eta)/(k\eta)$, the $C^{TT(t)}_l$ becomes
\beq
C^{TT(t)}_l= \frac{\pi\,r\,A_s}{4}\frac{(l+2)!}{(l-2)!}\int_{0}^{\infty} \frac{du}{ u}
\left(\int_{\xi_r}^{1}d\xi\frac{\partial }{\partial\xi}\left(\frac{3j_1(u\xi)}{u\xi}\right)
\frac{j_{l}[(1-\xi)u]}{(1-\xi)^2u^2}\right)^{2}~.\label{cltt0c}
\eeq
With $r=0.1$ and $A_s=2.2\times 10^{-9}$ we numerically integrate \eqref{cltt0c} and compare it with the results of CAMB CMB code \cite{Lewis:1999bs}. Here we set the Planck 2013 best fit parameters \cite{Ade:2013zuv} in CAMB. We also do not consider the effects of reionization on the temperature and polarization anisotropies and the effects of neutrino on the amplitude of tensor perturbations. Therefore we switch off both effects in the CAMB program. From Fig. 2.(a), we see a fair agreement between results of CAMB and the analytical results of \eqref{cltt0c} for $l<50$.

\begin{figure}
    \centering
        \subfigure[  ]
    {
        \includegraphics[width=3.3in]{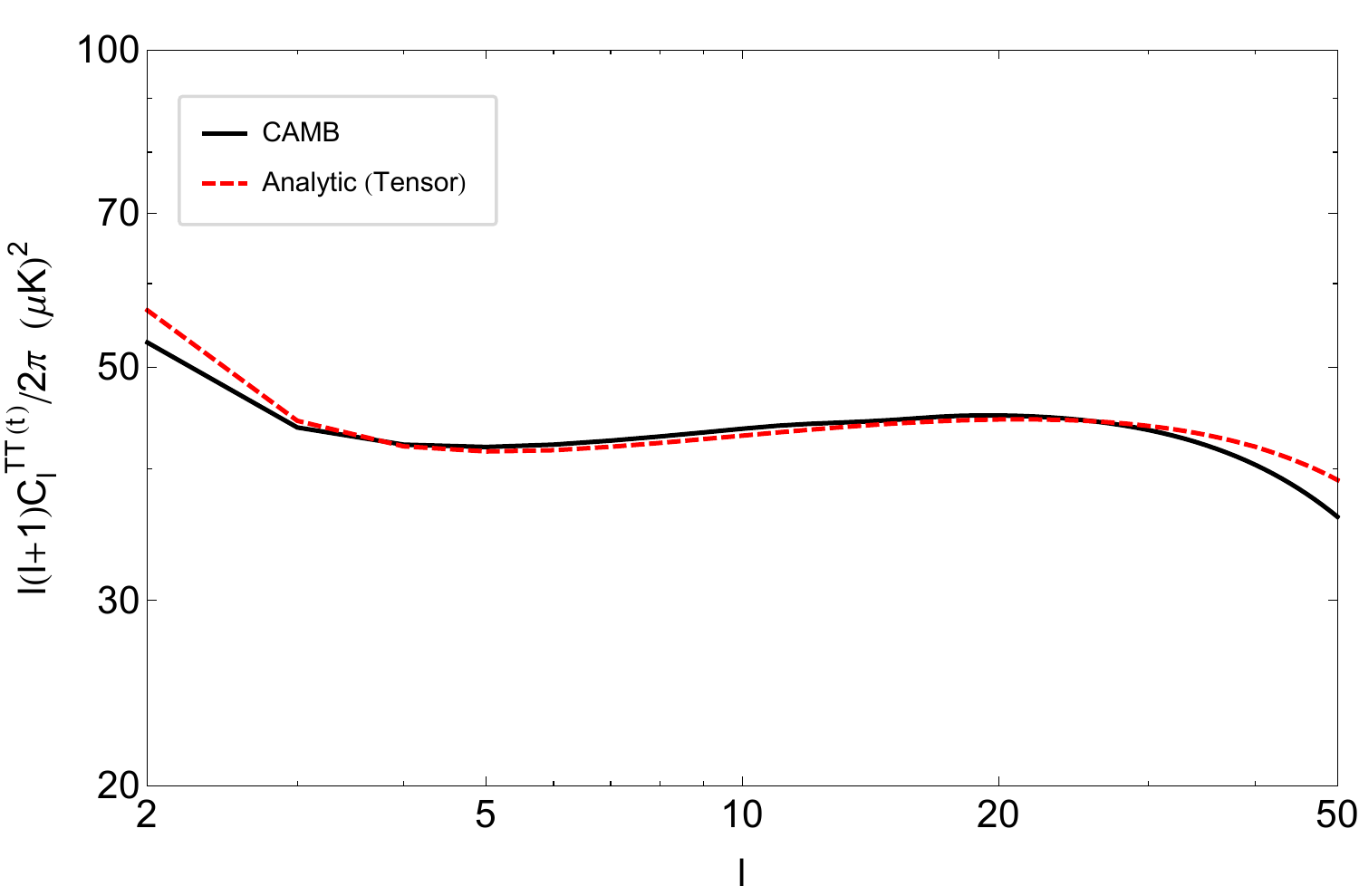}
        \label{fig:second_sub}
    }
       \subfigure[]
    {
        \includegraphics[width=3.33in]{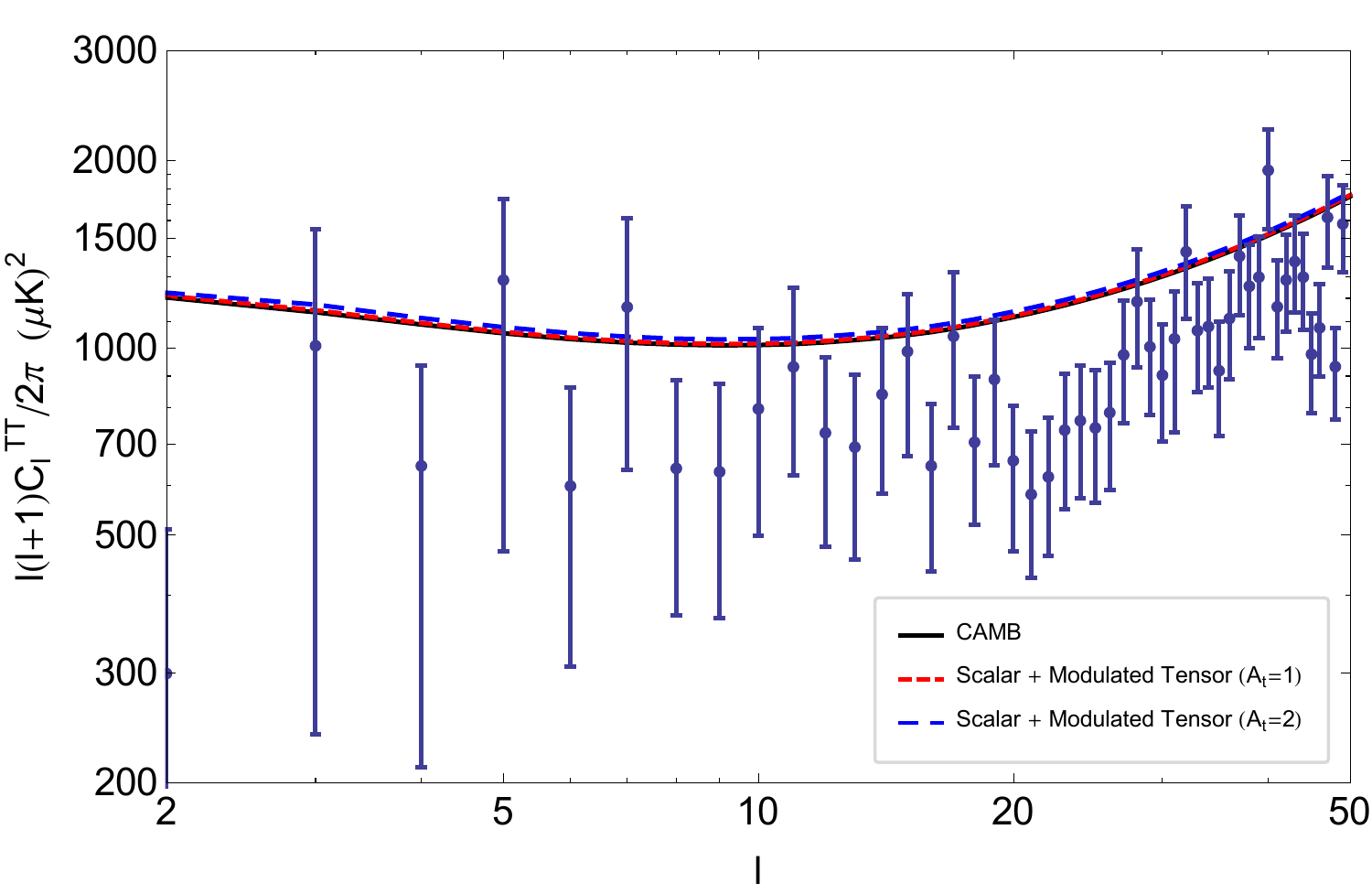}
        \label{fig:third_sub}
    }
    \caption{ The comparison of the full numerical results of CAMB and the analytical results derived in the text for (a) $C^{TT(t)}_{l}$ and (b) total angular power spectrum $C^{TT}_{l}$. Points show the Planck 2013 data.  }
    \label{fig:sample_subfigures}
\end{figure}

We want to extend the calculations leading to Eq.\eqref{cltt0c} to the case in which the tensor modes are modulated. To proceed, we first replace $h^{TT}_{ij}$ with $h^{TT}_{ij}(1+A_t\cos\theta)$ and then divide the multipoles into two parts $a^{t}_{lm}+\delta a^{t}_{lm}$ such that the second part contains the $A_t\,h^{TT}_{ij}\cos\theta$. The method of calculation of $\delta a^{t}_{lm}$ is the same as described above for the $a^{t}_{lm}$ but rather more complex, so we do not present all details. After some straightforward calculations we arrive at the following expression for  $\delta a^{t}_{lm}$
\bea
\delta a^{t}_{l\pm 2}&=&A_{t}\left[h_{(i)}^{(+)}\mp ih_{(i)}^{(\times)}\right]\frac{\pi}{2}\sqrt{\frac{2l+1}{4\pi}\frac{(l-2)!}{(l+2)!}}\int_{-1}^{1} dx x \sum_{l'=0}^{\infty}(2l'+1)i^{l'}\nonumber \\
&&\:\:\:\:\:\:\:\:\:\:\:\:\:\:\:\:\:\:\:\:\:\:\:\:\times
\left(2lxP_{l-1}-l(l+1)P_l+l(l-1)x^2P_{l}\right)P_{l'}(x)\int_{\eta_r}^{\eta_0}d\eta\frac{\partial H}{\partial\eta}j_{l'}[(\eta_0-\eta)k]~,
\eea
where $x=\cos\theta$. The integration over the $x$ variable can be performed by using again the recurrence and orthogonality relations of Legendre polynomials \eqref{legendre}. We find
 \bea
\delta a^{t}_{l\pm 2}&=&A_{t}\frac{\pi i^{l+1}}{2l+1}\left[h_{(i)}^{(+)}\mp ih_{(i)}^{(\times)}\right]\sqrt{\frac{2l+1}{4\pi}\frac{(l+2)!}{(l-2)!}}\int_{\eta_r}^{\eta_0}d\eta\frac{\partial H}{\partial\eta}\left[ \frac{l-2}{(2l-1)(2l-3)}j_{l-3}[(\eta_0-\eta)k] \right. \nonumber \\ &&\left. \:\:\:\:\:\:\:\:\:\:\:\:\:\:\:\:\:\:\:\:\:\:\:\:+\frac{l-3}{(2l-3)(2l+3)}j_{l-1}[(\eta_0-\eta)k] -\frac{l+4}{(2l-1)(2l+5)}j_{l+1}[(\eta_0-\eta)k]\right. \nonumber \\ &&\left. \:\:\:\:\:\:\:\:\:\:\:\:\:\:\:\:\:\:\:\:\:\:\:\: -\frac{l+3}{(2l+3)(2l+5)}j_{l+3}[(\eta_0-\eta)k]
\right]~.\label{dela}
\eea
Using \eqref{cltt} one can also define
\bea
\delta C^{TT(t)}_l=\frac{1}{2l+1}\int d^{3}k\left<\mid \delta a^{t}_{l2}(\mathbf{k})\mid^{2}+\mid \delta a^{t}_{l-2}(\mathbf{k})\mid^{2}\right>~,
\eea
where this expresses the contribution of dipole modulation in the tensor angular power spectrum. Therefore, using \eqref{dela} we get
\beq
 \delta C^{TT(t)}_{l}= A_t^2\,\delta^{(1)}_l~,\label{deltaclt}
 \eeq
where
\bea
\delta^{(1)}_l&=&\frac{\pi\,r\,A_s}{4}\frac{(l+2)!}{(l-2)!}\int_{0}^{\infty} \frac{du}{ u}\left\{
\int_{\xi_r}^{1}d\xi\frac{\partial }{\partial\xi}\left(\frac{3j_1(u\xi)}{u\xi}\right)\left[\frac{l-2}{(2l+1)(2l-1)(2l-3)}j_{l-3}[(1-\xi)u]\right. \right. \nonumber \\ &&\left.\left.\:\:\:\:\:\:\:\:\:\:\:\:\:\:\:\:\:\:\:\:\:\:\:\:\:+\frac{l-3}{(2l+1)(2l-3)(2l+3)}j_{l-1}[(1-\xi)u] -\frac{l+4}{(2l+1)(2l-1)(2l+5)}j_{l+1}[(1-\xi)u] \right. \right. \nonumber \\ &&\:\:\:\:\:\:\:\:\:\:\:\:\:\:\:\:\:\:\:\:\:\:\:\:\:\left.\left.-\frac{l+3}{(2l+1)(2l+3)(2l+5)}j_{l+3}[(1-\xi)u]\right]\right\}^2~,
\label{delta2}
\eea
so that the tensor angular spectrum is given by $ C^{TT(t)}_l+A_t^2\,\delta^{(1)}_l$. Setting $A_t=0$ gives rise to the unmodulated case. Here, we have not considered the dipole modulation in the scalar perturbations. Hence the total angular power spectrum is
\beq
C^{TT}_l=C^{TT(\zeta)}_l+C^{TT(t)}_l+A_t^2\,\delta^{(1)}_l
\eeq
We keep the curvature perturbation $\zeta$ unmodulated hence the $C^{TT(\zeta)}_l$ spectrum is calculated using the CAMB code. The $C^{TT(t)}_l$ and the $A_t^2\,\delta^{(1)}_l$ factors are also given by numerically integrating the equations \eqref{cltt0c} and \eqref{deltaclt}. Then we combine the $C^{TT(\zeta)}_l$ given by CAMB with the $C^{TT(t)}_l+A_t^2\,\delta^{(1)}_l$ given by equations \eqref{cltt0c} and \eqref{deltaclt} and obtain the total angular power spectrum $C^{TT}_l$.
In Fig. 2(b) we have shown the resulting $C^{TT}_l$ with $A_t=1$ and $2$. They have been compared with the total angular power spectrum derived by CAMB.
As the curves depicted in Fig 2(b) clearly manifest the dipole modulation in tensor perturbations with $A_t\sim 1$ does not make a considerable contribution to the $C^{TT}_l$. For $l\sim 10$ we see a small deviation from the non modulated case which falls down for $l>10$. Note that these effects are one order of magnitude smaller in the $C^{EE}_{l}$ and $C^{TE}_{l}$ spectra. As briefly discussed in the introduction, the amplitude of the modulation measured by Planck is $A= 0.078$. For small $l$, we find the corresponding value of $A_t$ required to produce this
amplitude as $A_t\sim A/\sqrt{r}\sim 0.1$ for $r\sim 0.1$.

We can also calculate the general two point correlator $C^{TT(t)}_{l_1m_1l_2m_2}$ \cite{Dvorkin:2007jp,Hanson,soda,Bartolo:2014hwa}. In order to find the nonzero elements of correlator we use the approach of \cite{soda}. In this method we assume that $\mathbf{k}$ makes the angle $\theta$ with the preferred direction. The transfer function is denoted by $\Delta_{l}(k,\eta)$ that can be calculated in terms of $H(k,\eta)$ and the spherical Bessel functions $j_{l}(k,\eta)$. After some straightforward calculations the multipoles $a^{t}_{lm}$ are generally given by \cite{soda}
\bea
a^{t}_{lm}=\int d^{3}k \Delta_{l}(k,\eta)\left[h_{(i)}^{+2}\:_{-2}Y^{\ast}_{lm}(\mathbf{n}_{k})+h_{(i)}^{-2}\:_{2}Y^{\ast}_{lm}(\mathbf{n}_{k})\right]~,
\eea
where $h_{(i)}^{\pm2}=h_{(i)}^{+}\mp ih_{(i)}^{\times}$ and $_{s}Y_{lm}$ are the spin-weighted spherical harmonics. The correlator $C^{TT(t)}_{l_1m_1l_2m_2}$ is defined as
\beq
C^{TT(t)}_{l_1m_1l_2m_2}=\left<a^{t}_{l_1m_1}a^{t\ast}_{l_2m_2}\right>~.
\eeq
We now assume that the preferred direction to be coincided with the $z$ direction and using \eqref{anisop1}, we find
\bea
\delta C^{TT(t)}_{l_1m_1l_2m_2} &\propto & \frac{rA_sA_t^2}{15}\sqrt{ \frac{4\pi}{(2l_1+1)(2l_2+1)}}\int \frac{dk}{k} \Delta_{l_1}(k,\eta) \Delta_{l_2}(k,\eta) \nonumber \\&&\times
\int d\mathbf{n}_{k}\left[Y_{20}(\mathbf{n}_{k})\:_{-2}Y^{\ast}_{l_1m_1}(\mathbf{n}_{k})\:_{-2}Y_{l_2m_2}(\mathbf{n}_{k})
+Y_{20}(\mathbf{n}_{k})\:_{2}Y^{\ast}_{l_1m_1}(\mathbf{n}_{k})\:_{2}Y_{l_2m_2}(\mathbf{n}_{k})\right]~,
\eea
and using the properties of the spin-weighted spherical harmonics one can write
\bea
\delta C^{TT(t)}_{l_1m_1l_2m_2} \propto  \frac{2rA_sA_t^2}{15}\sqrt{ \frac{4\pi}{(2l_1+1)(2l_2+1)}}\int \frac{dk}{k} \Delta_{l_1}(k,\eta) \Delta_{l_2}(k,\eta)
\int d\mathbf{n}_{k} Y_{20}(\mathbf{n}_{k})\:_{2}Y_{l_1m_1}(\mathbf{n}_{k})\:_{2}Y_{l_2m_2}(\mathbf{n}_{k}) ~.
\eea
The integral over $\mathbf{n}_{k}$ is calculated using the Gaunt integral formula and the solution is given in terms of the Wigner 3j-symbols
\bea
\delta C^{TT(t)}_{l_1m_1l_2m_2} \propto  \frac{2\sqrt{3}rA_sA_t^2}{15}\int \frac{dk}{k} \Delta_{l_1}(k,\eta) \Delta_{l_2}(k,\eta)
\left(
                                \begin{array}{ccc}
                                  2 & l_1 & l_2 \\
                                  0 & m_1 & m_2 \\
                                \end{array}
                              \right)\left(
                                \begin{array}{ccc}
                                  2 & l_1 & l_2 \\
                                  0 & 2 & 2 \\
                                \end{array}
                              \right)~.
\eea
By taking into account the selection rules, the last Wigner 3j-symbol is zero unless $l_1+l_2+2=\textrm{even}$ and $|l_1-2|\leq l_2\leq l_1+2$. These conditions allows the nonzero $\delta C^{TT(t)}_{l_1m_1l_2m_2}$ for $l_2=l_1,\,l_1\pm 2$.

\section{The effects of dipole modulation on $C_l^{BB}$}

The polarization of CMB is quantified by Stokes parameters $Q(\mathbf{n})$ and $U(\mathbf{n})$ measured as a function of position on the sky. It is known that the combination $Q(\mathbf{n})\pm i\,U(\mathbf{n})$ transforms like a spin-2 variable under rotation. Hence expanding this combination in spin weighted spherical harmonics, $\, _{\pm2}Y_{lm}$, gives
\beq
(Q\pm i\,U)(\mathbf{n})=\sum_{l=2}^{\infty}\sum_{m=-l}^{+l}a_{lm}^{\pm 2}\, _{\pm2}Y_{lm}(\mathbf{n})~.
\eeq
This help us to define two E- and B-modes by linear combinations of coefficients $a_{lm}^{\pm 2}$
\beq
a^{E}_{lm}=-\frac{1}{2}\left(a_{lm}^{+ 2}+a_{lm}^{- 2}\right)\:\:\:\:\:\:\textrm{and}\:\:\:\:\:\:a^{B}_{lm}=\frac{i}{2}\left(a_{lm}^{+ 2}-a_{lm}^{- 2}\right)~,
\eeq
where E-modes are invariant under the parity transformations while B-modes change sign. Usually the full sky polarization map of CMB is decomposed into E-mode and B-mode \cite{Kamionkowski:1996ks,Zaldarriaga:1996xe}. Physically the E-mode polarization is generated by scalar and tensor perturbations. It can be shown that the B-mode is just generated by the tensor perturbation. Therefore, the B-mode can probe the primordial gravitational wave. Any dipole modulation in tensor modes can imprint on both E mode and B-mode. However, we expect larger effects on the B-mode. In order to calculate the $a^{E,B}_{lm}$ multipoles as it is convenient we define the polarization matrix in terms of Stokes Parameters
\bea
\mathcal{P}_{ab}(\mathbf{n})&=&\int d^{3}k\mathcal{P}_{ab}(\mathbf{k},\mathbf{n})\nonumber \\&&\!\!\!\!\!\!\!\!=\frac{1}{2}\left(
                                          \begin{array}{cc}
                                            Q(\mathbf{n}) & -U(\mathbf{n})\sin\theta \\
                                            -U(\mathbf{n})\sin\theta & -Q(\mathbf{n})\sin^2\theta \\
                                          \end{array}
                                        \right)~.
\eea
Hence the coefficient $a^{E,B}_{lm}$ are given by
\beq
a^{E,B}_{lm}=-\int d\mathbf{n}Y^{(E,B)\ast}_{lm,ab}(\mathbf{n})\mathcal{P}^{ab}(\mathbf{n})~,\label{almEB}
\eeq
where
\beq
Y^{(B)}_{lm,ab}(\mathbf{n})=\sqrt{\frac{(l-2)!}{2(l+2)!}}\left(
                                                              \begin{array}{cc}
                                                                -X_{lm}(\mathbf{n}) & W_{lm}(\mathbf{n})\sin\theta \\
                                                                W_{lm}(\mathbf{n})\sin\theta & X_{lm}(\mathbf{n})\sin^2\theta \\
                                                              \end{array}
                                                            \right)~,
\eeq
with the auxiliary functions $X_{lm}$ and $W_{lm}$ constructed as
\bea
 W_{lm}(\mathbf{n})=\left(2\frac{\partial^{2}}{\partial\theta^{2}}+l(l+1)\right)Y_{lm}(\mathbf{n})~, \label{wlm}\\
 X_{lm}(\mathbf{n})=\frac{2im}{\sin\theta}\left(\frac{\partial}{\partial\theta}-\frac{\cos\theta}{\sin\theta}\right)Y_{lm}(\mathbf{n})~.\label{xlm}
\eea

The parameters of polarization matrix and also the CMB angular spectra are mostly derived by a hierarchy of Boltzmann equations \cite{Zaldarriaga:1996xe,Hu:1997mn}. Instead, we take an analytic approach proposed in \cite{Kamionkowski:1996ks,Gorbunov:2011zzc} to study the CMB polarization. We compare our results with the methods implemented in the Boltzmann code CAMB to check the analytical method. We then extend the analytical calculation to include the modulation in the tensor modes. The Fourier transformation of polarization matrix $\mathcal{P}_{ab}(\mathbf{k},\mathbf{n})$ for tensor perturbations is analytically given by the following matrix \cite{Gorbunov:2011zzc}

\bea
\mathcal{P}^{t}_{ab}(\mathbf{k},\mathbf{n})&=&\frac{\Delta\eta_r}{10}\frac{\partial H}{\partial\eta}\,e^{ik(\eta_0-\eta_r)\cos\theta}\nonumber \\&& \times\left(
                                                              \begin{array}{cc}
                                                                -\left(1+\cos^2\theta\right)\left\{\cos 2\phi\, h_{(i)}^{+}+\sin 2\phi \,h_{(i)}^{\times}\right\} & \sin 2\theta\left\{\sin 2\phi \,h_{(i)}^{+}+\cos 2\phi\, h_{(i)}^{\times}\right\} \\
                                                                \sin 2\theta\left\{\sin 2\phi\, h_{(i)}^{+}+\cos 2\phi \,h_{(i)}^{\times}\right\}&  \left(1+\cos^2\theta\right)\left\{\cos 2\phi \, h_{(i)}^{+}+\sin 2\phi \,h_{(i)}^{\times}\right\} \\
                                                              \end{array}
                                                            \right)~,\label{polarmatrix}
\eea
where $H(k,\eta)$ is again the transfer function for tensor modes and $\Delta\eta_r$ is the thickness of the last scattering sphere. Note that in this expresion we have not considered the gravitational lensing and also the reionization effect.
One can easily show that in the scalar perturbations case the off diagonal components of polarization tensor vanish. However, for the tensor perturbations, the new terms supplied by gravity waves result in non-vanishing values for the Stokes parameter $U$ has a principal role in generating the B mode polarization. Now after computing the polarization matrix \eqref{polarmatrix} one can find the coefficients $a^{B}_{lm}$ by using the relation \eqref{almEB}. We defer the details of calculation to the Appendix. By using the results presented in the Appendix we can evaluate the parity independent angular power spectra $C_{l}^{BB}$ as follows
\bea
C^{BB}_l=\frac{2\pi}{25}rA_s\Delta\eta_r^{2}\int_{0}^{\infty}\frac{dk}{k}\left(\frac{\partial H(k,\eta_r)}{\partial\eta}\right)^2
\left[\frac{l+2}{2l+1}j_{l-1}(k\eta_0)-\frac{l-1}{2l+1}j_{l+1}(k\eta_0)\right]^2~,\label{ClBB}
\eea
The transfer function is computed at the time $\eta=\eta_r$. As we discussed in section II at this time one can approximate the transfer function by $H(k,\eta_r)=\sin(k\eta_{r})/(k\eta_{r})$. Changing the integration variables to $\xi$ and $u$ we find
\bea
C^{BB}_l=\frac{2\pi}{25}rA_s\Delta\xi_r^{2}\int_{0}^{\infty}\frac{du}{u}\left(\cos(u\xi_{r})-\frac{\sin(u\xi_{r})}{u\xi_{r}}\right)^2
\left[\frac{l+2}{2l+1}j_{l-1}(u)-\frac{l-1}{2l+1}j_{l+1}(u)\right]^2~,\label{ClBB2}
\eea
We have actually found that the analytical expression \eqref{ClBB2} has a good agreement with the $C_{l}^{BB}$ calculated by CAMB with $\Delta\xi_r=0.028$ at $l<100$. In Fig. 3 we see this agreement with $r=0.1$ and $A_s=2.2\times10^{-9}$. At $l<10$ the $C_{l}^{BB}$ curve grows up while the analytical curve displays an opposite behavior. This is due to impact of reionization on the CMB which we have not considered in this work.

\begin{figure}\vspace{.5in}
      \includegraphics[width=4in]{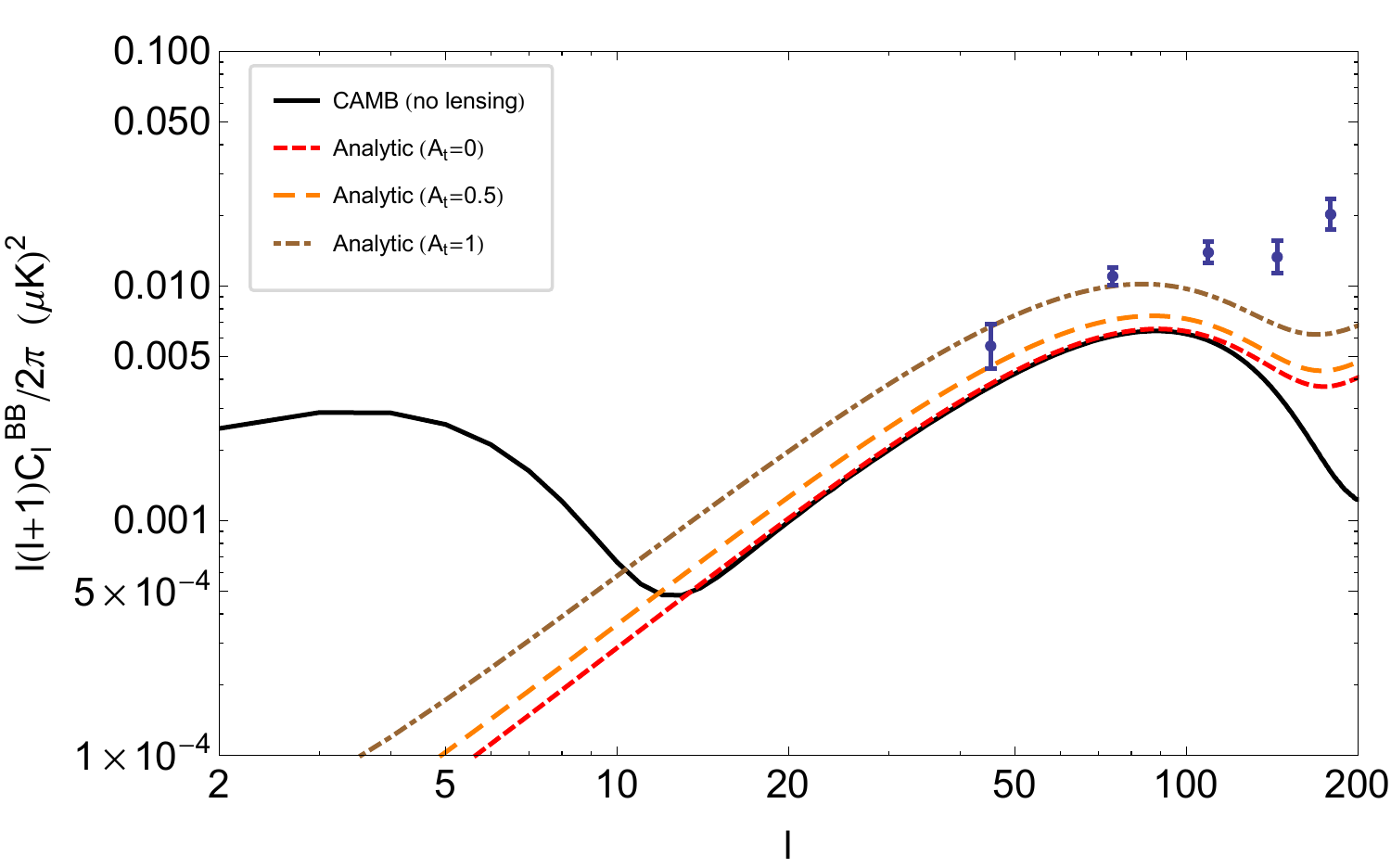}
    \caption{The comparison of the BB angular power spectrum calculated by CAMB with $r=0.1$, $A_s=2.2\times 10^{-9}$, $n_t=0$ with the analytical results derived in the text. Points show the BICEP2/Keck Array data. }
    \label{powerchotic}
\end{figure}

We now consider the effects of the modulation in tensor modes on the angular power spectra of CMB. Recall that to derive the multipole coefficients we need to perform the integration over all angles $\theta$. As we discussed the modulation contributes the new factor $(1+A_{t}\cos\theta)$ in front of the integrand. We therefore separate the multipole coefficients into $a^{B}_{lm}+\delta a^{B}_{lm}$ where the $\delta a^{B}_{lm}$ are those containing the $A_{t}\cos\theta$ term. The details of the calculation of $\delta a^{B}_{lm}$ coefficients are presented in Appendix. Using these results one can derive

\bea
\delta C^{BB}_l=A_t^{2}\delta^{B}_{l}~,
\eea
where
\bea
\delta^{B}_{l}=\frac{2\pi}{25}rA_s\Delta\xi_r^{2}\int_{0}^{\infty}\frac{du}{u}\left(\cos(u\xi_{r})-\frac{\sin(u\xi_{r})}{u\xi_{r}}\right)^2
&&\!\!\!\!\!\!\!\left[\frac{(l-1)(l+2)}{(2l+1)(2l-3)}j_{l-2}(u)-\frac{2l^2+2l+1}{(2l-1)(2l+3)}
j_{l}(u)\right. \nonumber \\ &&\left.
\:\:\:\:+\frac{(l-1)(l+2)}{(2l+1)(2l+5)}j_{l+2}(u)\right]^2~,\label{deltaClBB2}
\eea
where we have changed the variables of integration to $\xi$ and $u$. By considering the modulation the total BB power spectrum will be
 \beq
 C^{BB}_l= C^{BB}_l+A_t^{2}\delta^{B}_{l}~.\label{clbb3}
\eeq
In Fig. 3 we have also plotted the total predicted BB power spectrum for $A_t=0.5$ and $A_t=1$. As we can see the $C^{BB}_l$ is shifted above due to the modulation term in \eqref{clbb3}.

As well as, the $C^{BB}_{l_1m_1l_2m_2}$ correlator can be calculated using the method discussed in previous section. Similar to the TT case, it can be shown that the BB correlation is nonzero for $l_2=l_1,\:l_1\pm2$.

\section{Conclusion }

In this work we have studied the imprints of dipole modulation in tensor modes on the $C^{XY}_l$ with $XY=TT$ and $BB$. The modulation of tensor modes can be due to a long wavelength scalar or tensor mode which is superhorizon during inflation. Here we have modulated the tensor mode by multiplying its amplitude by a modulated factor like $\left(1+\sin(\mathbf{k}\cdot \mathbf{x}_{\textrm{lss}})\right)$. The angular power spectra of CMB have been analytically computed in the presence of the modulation factor. With a modulation in tensor modes one can see a larger modification in the $C^{BB}_l$. We also showed that the TT and BB correlators are allowed for the configurations $l_2=l_1,\:l_\pm 2$.
 The future detection of gravitational waves can constraint the amplitude of modulation. However this task needs a comprehensive study of the effects of modulation in tensor modes on the CMB temperature and polarization anisotropies. Here we have not considered the reionization and lensing effects. Either of these phenomena can change the simplified picture studied in this work.

\section*{\small Acknowledgment}

The author would like to thank H. Firouzjahi, A. Abolhasani, R. Emami, M. H. Namjoo for various comments and useful discussions. The author would like to thank H. Firouzjahi and R. Emami for collaboration in the early stages of this project.

\appendix

\section{The calculation of $C_l^{BB}$}

Here we calculate analytical expression for the BB polarization spectra of tensor perturbations. To this purpose, we need to know the contribution of tensor modes to the polarization multipole coefficients $a^{B}_{lm}$. Using the polarization matrix elements \eqref{polarmatrix} one can write
\beq
a^{B}_{lm,(+)}(\mathbf{k})=-\frac{\Delta\eta_{r}}{5}\frac{\partial H}{\partial\eta}\,h^{+}_{(i)}\sqrt{\frac{(l-2)!}{2(l+2)!}}\int d\mathbf{n}\,e^{i(\eta_0-\eta_r)k\cos\theta}\left[X^{\ast}_{lm}(1+\cos^2\theta)\cos 2\phi+2\,W^{\ast}_{lm}\cos\theta\sin 2\phi\right]~,\label{ablm1}
\eeq
and
\beq
a^{B}_{lm,(\times)}(\mathbf{k})=-\frac{\Delta\eta_{r}}{5}\frac{\partial H}{\partial\eta}\,h^{\times}_{(i)}\sqrt{\frac{(l-2)!}{2(l+2)!}}\int d\mathbf{n}\,e^{i(\eta_0-\eta_r)k\cos\theta}\left[X^{\ast}_{lm}(1+\cos^2\theta)\sin 2\phi-2\,W^{\ast}_{lm}\cos\theta\cos 2\phi\right]~,\label{ablm2}
\eeq
Inserting the \eqref{wlm} and \eqref{xlm} into equations \eqref{ablm1} and \eqref{ablm2}, changing the variable of integration from $\theta$ to $x$ and integrating over the azimuthal angle $\phi$ we get
\beq
a_{l2}^{B+}=\frac{2\pi i}{5\sqrt{2}}\,h_{(i)}^{(+)}\frac{\partial H}{\partial\eta}\sqrt{\frac{2l+1}{4\pi}}\Delta\eta_{r}\int_{-1}^{1}d x\,e^{ikx(\eta_0-\eta_r)}\left[\frac{l+2}{2l+1}P_{l-1}(x)-\frac{l-1}{2l+1}P_{l+1}(x)\right]~,\label{al2b}
\eeq
and as well as $a_{l2}^{B+}=-\,a_{l-2}^{B+}$, $a_{l2}^{B\times}=i\,a_{l2}^{B+}$ and $a_{l-2}^{B\times}=-i\,a_{l-2}^{B+}$. From the series expansion of plane wave in terms of Legendre polynomials \eqref{planew} one can find

\beq
a_{l2}^{B+}=\frac{4\pi i^{l}}{5\sqrt{2}}\,h_{(i)}^{(+)}\frac{\partial H}{\partial\eta}\sqrt{\frac{2l+1}{4\pi}}\Delta\eta_{r}\left[\frac{l+2}{2l+1}j_{l-1}(k(\eta_0-\eta_r))-\frac{l-1}{2l+1}j_{l+1}(k(\eta_0-\eta_r))\right]~.
\eeq
The angular power spectrum $C_l^{BB}$ is given by
\bea
C_{l}^{BB}&=&\frac{1}{2l+1}\int d^{3}k\left<a_{l2}^{B+}a_{l2}^{B+\ast}+a_{l2}^{B\times}a_{l2}^{B\times\ast}+a_{l-2}^{B+}a_{l-2}^{B+\ast}+a_{l-2}^{B\times}a_{l-2}^{B\times\ast}\right>
\nonumber \\&& \!\!\!\!\!\!\!\!=\frac{4}{2l+1}\int d^{3}k\left<a_{l2}^{B+}a_{l2}^{B+\ast}\right>~.
\eea
Therefore by inserting the $a_{l2}^{B+}$ we find
\bea
C^{BB}_l=\frac{2\pi}{ 25}\Delta\eta_r^{2}\int_{0}^{\infty}\frac{dk}{k}\mathcal{P}_{t}\left(\frac{\partial H(k,\eta_r)}{\partial\eta}\right)^2
\left[\frac{l+2}{2l+1}j_{l-1}(k\eta_0)-\frac{l-1}{2l+1}j_{l+1}(k\eta_0)\right]^2~.
\eea
In the case of modulation in tensor mode we have
\beq
\delta a_{l2}^{B+}=A_t\frac{2\pi i}{5\sqrt{2}}\,h_{(i)}^{(+)}\frac{\partial H}{\partial\eta}\sqrt{\frac{2l+1}{4\pi}}\Delta\eta_{r}\int_{-1}^{1}d x\,e^{ikx(\eta_0-\eta_r)}\,x\,\left[\frac{l+2}{2l+1}P_{l-1}(x)-\frac{l-1}{2l+1}P_{l+1}(x)\right]~,\label{al2b1}
\eeq
and some calculations yield

\bea
\delta a_{l2}^{B+}=-A_t\frac{4\pi i^{l+1}}{5\sqrt{2}}\,h_{(i)}^{(+)}\frac{\partial H}{\partial\eta}\sqrt{\frac{2l+1}{4\pi}}\Delta\eta_{r}&&\!\!\!\!\!\!\!\left[\frac{(l-1)(l+2)}{(2l+1)(2l-3)}j_{l-2}(k(\eta_0-\eta_r))-\frac{2l^2+2l+1}{(2l-1)(2l+3)}
j_{l}(k(\eta_0-\eta_r))\right. \nonumber \\ &&\left.
\:\:\:\:+\frac{(l-1)(l+2)}{(2l+1)(2l+5)}j_{l+2}(k(\eta_0-\eta_r))\right]~.
\eea
We define
\bea
\delta C_{l}^{BB}=\frac{4}{2l+1}\int d^{3}k\left<\delta a_{l2}^{B+}\delta a_{l2}^{B+\ast}\right>~.
\eea

Therefore the $\delta C_{l}^{BB}$ is found to be
\bea
\delta C^{BB}_l=\frac{2\pi}{ 25}A_t^{2}\Delta\eta_r^{2}\int_{0}^{\infty}\frac{dk}{k}\mathcal{P}_{t}\left(\frac{\partial H(k,\eta_r)}{\partial\eta}\right)^2
&&\!\!\!\!\!\!\!\left[\frac{(l-1)(l+2)}{(2l+1)(2l-3)}j_{l-2}[k(\eta_0)]-\frac{2l^2+2l+1}{(2l-1)(2l+3)}
j_{l}[k(\eta_0)]\right. \nonumber \\ &&\left.
\:\:\:\:+\frac{(l-1)(l+2)}{(2l+1)(2l+5)}j_{l+2}[k(\eta_0)]\right]^2~.
\eea




\end{document}